\documentclass[a4paper,11pt]{article}
\pdfoutput=1

\usepackage{jinstpub}
\usepackage{lineno}
\graphicspath{{figures/}}
\usepackage{subfig}

\title{Status and Prospects of the PandaX-III Experiment}

\author[a]{Wenming Zhang,}
\author[a]{Heng Lin,}
\author[a]{Yuanchun Liu,}
\author[a]{Ke Han,}
\author[a]{Kaixiang Ni,}
\author[a,b,*]{Shaobo Wang,\note[*]{Corresponding author.}}
\author[a]{Wenchang Zhai}
\author{(On behalf of PandaX-III collaboration)}

\affiliation[a]{INPAC and School of Physics and Astronomy, Shanghai Jiao Tong University, MOE Key Lab for Particle Physics, Astrophysics and Cosmology, Shanghai Key Laboratory for Particle Physics and Cosmology, Shanghai 200240, China}
\affiliation[b]{SPEIT (SJTU-Paris Elite Institute of Technology), Shanghai Jiao Tong University, Shanghai 200240, China}

\emailAdd{shaobo.wang@sjtu.edu.cn}

\abstract{
The PandaX-III experiment searches the neutrinoless double beta decay of $^{136}$Xe with a high-pressure xenon gaseous time projection chamber~(TPC). Thermal-bonding Micromegas modules are used for charge collection. Benefitting from the excellent energy resolution and imaging capability, the background rate can be significantly suppressed through the topological information of events. The technology is successfully demonstrated by a prototype detector. The final detector has been constructed. In this paper, we will report the status of the PandaX-III experiment, including the construction and commissioning of the final detector, and the Micromegas-based TPC performance test in the prototype detector.

}


\begin{document}
\maketitle
\flushbottom

\section{Introduction}
\label{sec:intro}
TPC (Time Projection Chamber) technology has been widely used in particle and nuclear physics experiments since its invention~\cite{Nygren:1978rx}. 
Nowadays, the application of TPC has been expanded to rare event searches, such as NLDBD (Neutrinoless Double Beta Decay) searches~\cite{NEXT:2023daz, EXO-200:2019rkq}, and dark matter direct detections~\cite{XENON:2023cxc, LZ:2022lsv, PandaX-4T:2021bab}. 
These rare event search experiments demand a very sensitive detector with low radioactivity, high energy resolution, and good imaging capabilities. 
A gaseous TPC records both energy depositions and trajectories of an event, providing powerful discrimination of the signal from the background for NLDBD searches~\cite{Qiao:2018edn, Galan:2019ake, Li:2021viv, Li:2022gpb, Xia:2022ddq}.

The PandaX-III experiment~\cite{Chen:2016qcd} uses a high-pressure gaseous TPC to search for NLDBD of $^{136}$Xe.
The active volume of the TPC is cylindrical with a height of 120.0~cm and a diameter of 160.0~cm.  It can hold about 140~kg of xenon (90\% enriched) at an operating pressure of 10 bar.
A tessellation of 52 ${\rm 20.0\times20.0~cm^{2}}$ thermal-bonding Micromegas modules~\cite{Feng:2019prv, Feng:2022jkd} are used for charge readout. 
The Micromegas with germanium (Ge) film-based resistive anode provides excellent energy resolutions and stabilities. Considering the low background, the electric field cage is designed with flexible PCBs (Printed Circuit Boards) connected with SMD (Surface Mount Devices) resistors, and supported by a low background acrylic barrel. All the technologies have been successfully demonstrated by a prototype detector. Currently, the construction of the detector and its sub-systems has been completed. We are doing the assembling and commissioning. 

In this paper, the overview of the PandaX-III experiment is first given in Section~\ref{sec:Overview}. Then the latest progress of the PandaX-III detector is presented in Section~\ref{sec:TPC}. The Micromegas test in the prototype detector is detailed in Section~\ref{sec:prototype}. Finally, we give the summary in Section~\ref{sec:outlook}.
%Charge signals recorded by Micromegas can reconstruct the electron tracks of xenon decay events, which helps suppress the background level.
%The detector will be operated in 10 bar Xe-(1${\rm \%}$)TMA (triethylamine) gas mixtures, and
%the TMA in xenon improves the detector performance attributed to the Penning effect~\cite{Cebrian:2012sp}.
%More details will be discussed below.

\section{PandaX-III experiment}
\label{sec:Overview}
\begin{figure}[htbp]
    \centering
    \includegraphics[width=.7\textwidth]{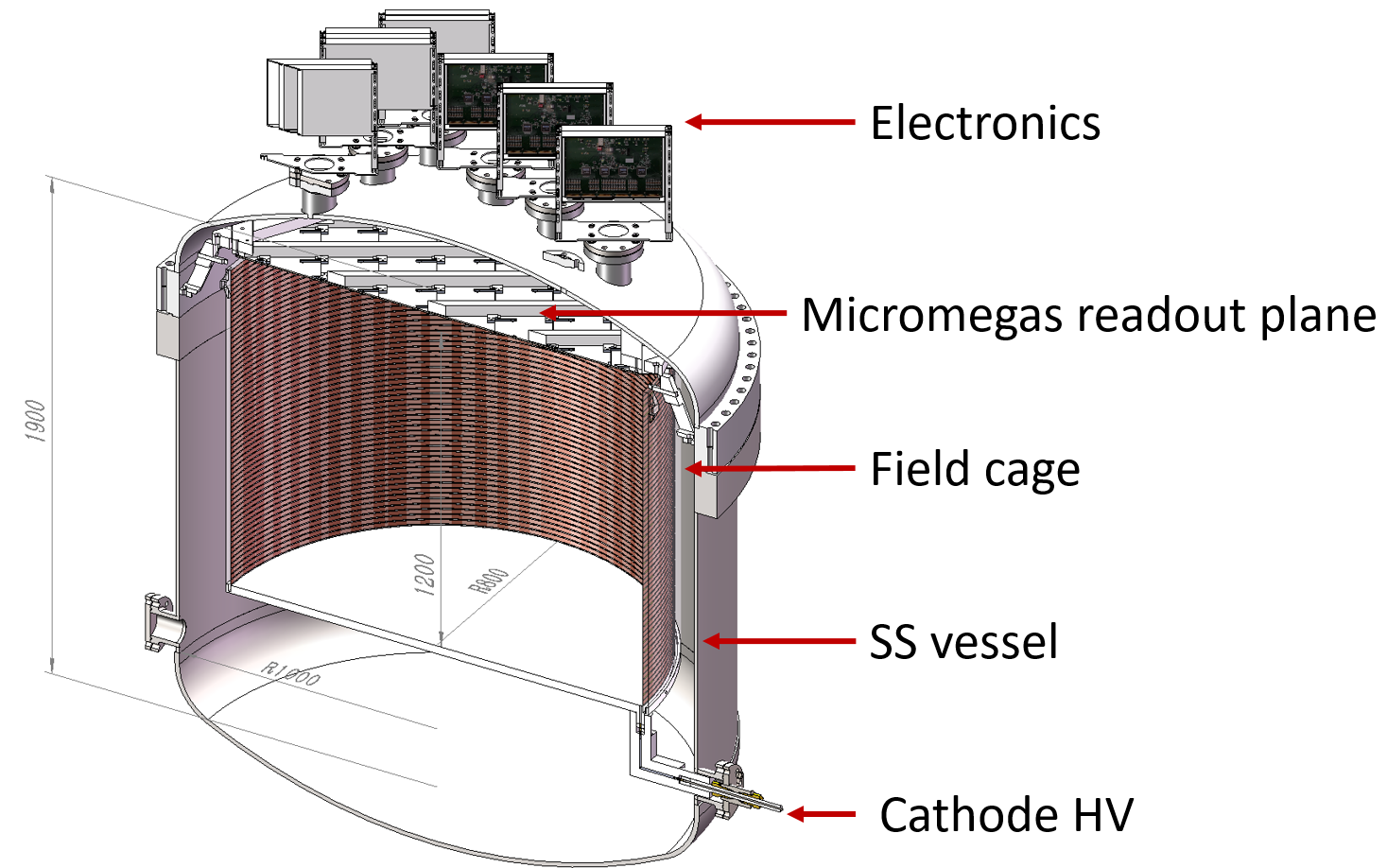}
    \caption{\label{fig:fullTPC_profile} Overview of the PandaX-III detector.}
\end{figure}
The PandaX-III experiment exploits a high-pressure gaseous TPC for NLDBD search.
The overview of the detector is shown in Fig.~\ref{fig:fullTPC_profile}.
The central part is a gaseous TPC consisting of a single-ended charge readout plane on the top, a cathode at the bottom, and an electric field shaping cage with a height of 120.0~cm and a diameter of 160.0~cm in between.
The charge readout plane with 52 Micromegas modules is suspended from the top flat flange of a high-pressure SS (stainless steel) vessel.
The working gas is 10 bar Xe-(1${\rm \%}$)TMA gas mixtures, so it contains about 140~kg 90\% enriched $^{136}$Xe inside the TPC. The detailed design is presented in~\cite{Wang:2020owr, Wang:2020csx}.

The TPC records both the energy deposition and three-dimensional tracks of events. 
The expected energy resolution is 3\% FWHM (Full Width at Half Maximum) at the Q-value (2458~keV). 
The track of MeV-scale electrons in 10 bar xenon is long enough (about 10.0 cm), therefore the topological information is powerful for signal-background discrimination~\cite{Qiao:2018edn, Galan:2019ake, Li:2021viv, Li:2022gpb, Xia:2022ddq}. The construction of the detector was completed.
The detector will be installed at the CJPL-II~\cite{Li:2014rca} after the performance test in Shanghai.
A background of 0.49 counts per year in the region of interest is expected with a signal-background discrimination algorithm applied, based on which
the exclusion sensitivity of PandaX-III for NLDBD half-life can reach 2.7$\times$10$^{26}$~yr~(90\% C.L.) with 5 years’ exposure~\cite{Li:2021viv}.

\section{PandaX-III detector}
\label{sec:TPC}
The construction of the PandaX-III detector is completed. A series of tests is ongoing at SJTU (Shanghai Jiaotong University).
%We paid special attention to the design and fabrication of individual parts during construction, as well as material selection to fulfill the requirement of vacuum and high pressure.
%We also conducted the radioactive contamination screening on low-background materials, such as stainless steel, which was used in the PandaX dark matter experiment~\cite{Tan:2016zwf}. 
The design and the performance test of the core detector components are described below.
\begin{figure}[tbp]
    \centering
    \includegraphics[width=.40\textwidth]{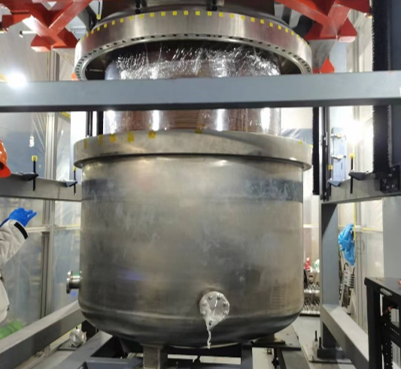}
    \includegraphics[width=.40\textwidth]{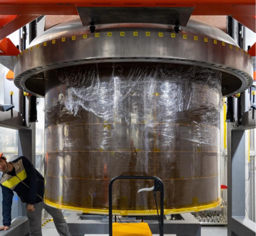}
    \qquad
    \includegraphics[width=.40\textwidth]{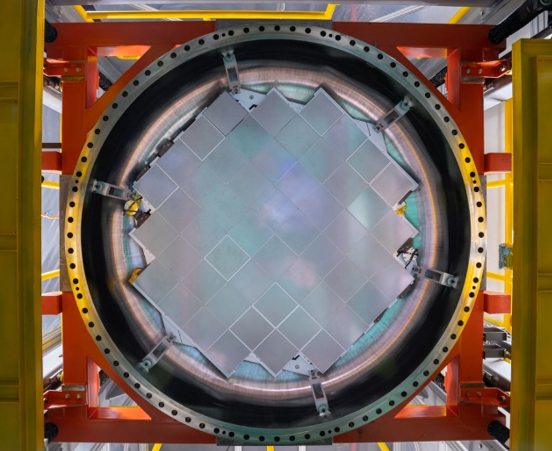}
    \includegraphics[width=.40\textwidth]{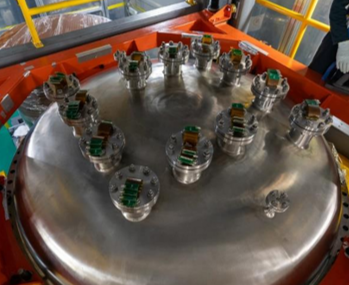}
    \caption{\label{fig:fullTPC_assembly} 
    Pictures of PandaX-III detector assembly. (Top-left) The high-pressure vessel. (Top-right) The electric field cage suspended from the top flange. (Bottom-left) The charge readout plane mounted with 52 Micromegas modules. (Bottom-right) The top dished flange features 12 DN-80 flanges, holding 52 Kapton extension cables.}
\end{figure}

\subsection{High-pressure SS vessel}
\label{sec: vessel}
The high-pressure vessel is made of low-background SS with pressurization up to 15 bar, as shown in Fig.~\ref{fig:fullTPC_assembly} (Top-left).
The vessel comprises a dished flange and a cylindrical barrel. The thickness of the dished flange is 1.6 to 1.8~cm with a height of 59.7~cm, which can provide enough mechanical strength.
The barrel is about 1.2~cm in thickness with a diameter diameter of 200.0~cm and a height of 127.9~cm.
The total height of the full detector is 232.7~cm.
84 M27 bolts are used to hold the top dished flange and barrel together. 

12 DN-80 flanges as shown in Fig.~\ref{fig:fullTPC_assembly} (Bottom-right) on the top dished flange are used to hold Kapton signal cables for the readout and high voltage supply of Micromegas.
The small DN-10 ports on the top dished flange are connected to a 1/4-inch SS pipe as a gas outlet.
Four DN-80 flanges are at the lower part of the barrel, one for high-voltage~(HV) feedthrough of the cathode, one for gas inlet, one for vacuum pumping, and one for backup.
All flanges are designed to withstand the required vacuum and high-pressure conditions, with expanded PTFE gaskets used for flange fittings. 
The installation and test of the main vessel have been completed, including the leakage check and high-pressure test. The pressure of vessel was monitored~\cite{Yan:2020fpm} and remained stable for over one month at 10 bar. 

%The tested leak rate is less than 0.01 ${\rm L \cdot bar/h}$, equivalent to 0.52~kg xenon leak per year.
%and the optimal vacuum reaches about 10 mPa at the pump end.

\subsection{Charge readout plane}
\label{sec:readout}

As shown in Fig.~\ref{fig:fullTPC_assembly} (Bottom-left), the charge readout plane consists of 52 Micromegas modules mounted on a specially designed aluminum holding frame.
%The Micromegas modules are fabricated with the so-called thermal-bonding technology, which makes an avalanche structure by pressing the stretched SS mesh directly onto the readout PCB using a hot rolling machine.
Each module has a square active area of ${\rm 20.0\times20.0~cm^2}$ with a flexible signal line (the tail). The amplification gap of Micromegas is 100~$\mu$m. 64 readout strips on each side (X and Y) with 3~mm pitch are embedded in the readout PCB. 
The signal cables, which are glued on the DN80 flanges, are designed to connect the Micromegas tail inside and the electronic system outside, as shown in Fig.~\ref{fig:fullTPC_assembly} (Bottom-right).
4 or 6 cables were glued on one flange, and it is estimated that about 3.4~g of xenon leaks per year at 10 bar for each flange. 
In total 52 cables glued on 12 flanges match with all Micromegas modules. 

\subsection{Electric field shaping cage}
\label{sec:FieldCage}

The electric field shaping cage is designed with flexible Kapton PCBs supported by a low-background acrylic barrel with a thickness of 4.8~cm. 
As shown in Fig.~\ref{fig:fullTPC_assembly} (Top-right), it is suspended from the top flange and defines a cylindrical drift volume of the TPC with a height of 120.0~cm and a diameter of 160.0~cm. 
Each piece of flexible PCB used for the field cage is a size of ${\rm 253.5\times21.5~cm^2}$.
Every 2 flexible PCBs are connected to form a ring, and 6 rings can cover the inner wall of the entire acrylic barrel.
120 SMD resistors of 1 Gohm are used to connect the copper strips, which are embedded in the flexible PCBs and have a width of 0.8 cm and a spacing of 0.2 cm.

\begin{figure}[tbp]
    \centering
    \includegraphics[width=5.8cm ,height=4.0cm]{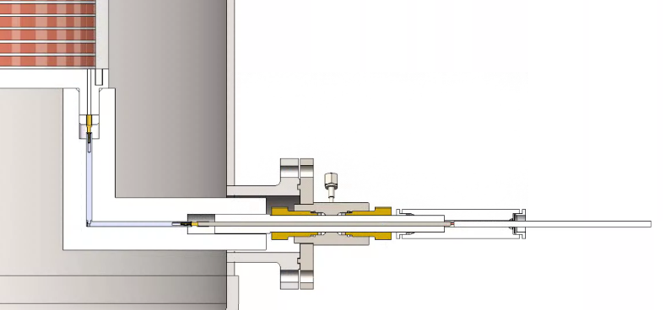}
    \includegraphics[width=.60\textwidth]{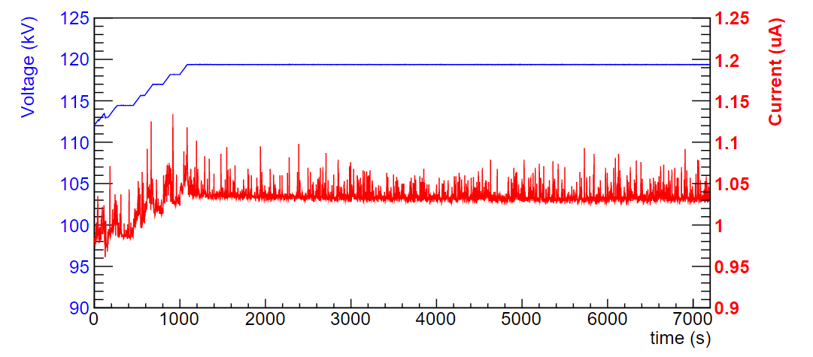}
    \caption{\label{fig:feedthrough_voltageTest_fullTPC} (Left) The design drawing of the L-shaped HV feedthrough connected to the cathode. (Right) The monitoring of voltage and current during the HV test of the electric field shaping cage.}
\end{figure}
 
The cathode plane on the bottom of the electric field cage is supported by an acrylic pedestal with a thickness of 17.0~cm for protection and insulation.
A negative HV is applied to the cathode through an L-shaped feedthrough in the DN-80 flange, as shown in Fig.~\ref{fig:feedthrough_voltageTest_fullTPC} (Left).
Therefore, ionized electrons drift upward in the drift volume with electric field lines pointing from top to bottom in the TPC.
The baseline design of the drift electric field is 1~kV/cm~\cite{Chen:2016qcd,Chaiyabin:2017mis}. 
The HV test was conducted in the air outside and inside the TPC.
Fig.~\ref{fig:feedthrough_voltageTest_fullTPC}~(Right) shows the result of the HV test with -120 kV applied to the cathode steadily in the air with a current of 1.03 ${\rm \mu A}$ without any sparks.

\subsection{Detector assembling and commissioning}
\begin{figure}[tbp]
    \centering
    \includegraphics[width=.48\textwidth]{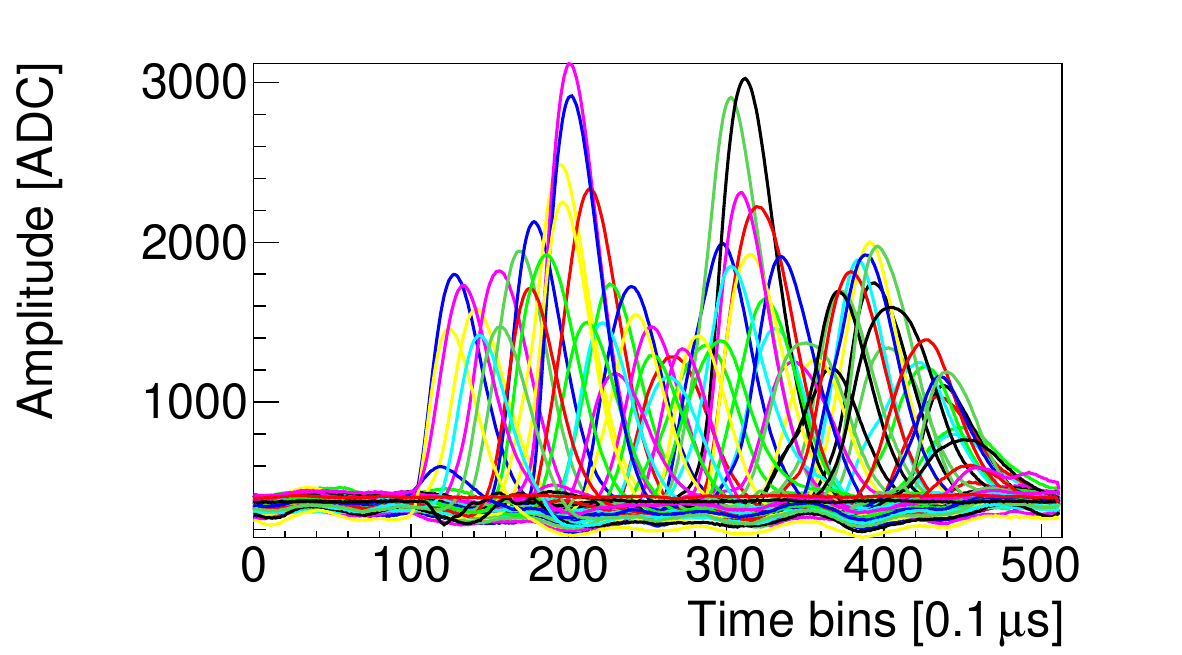}
    \includegraphics[width=.48\textwidth]{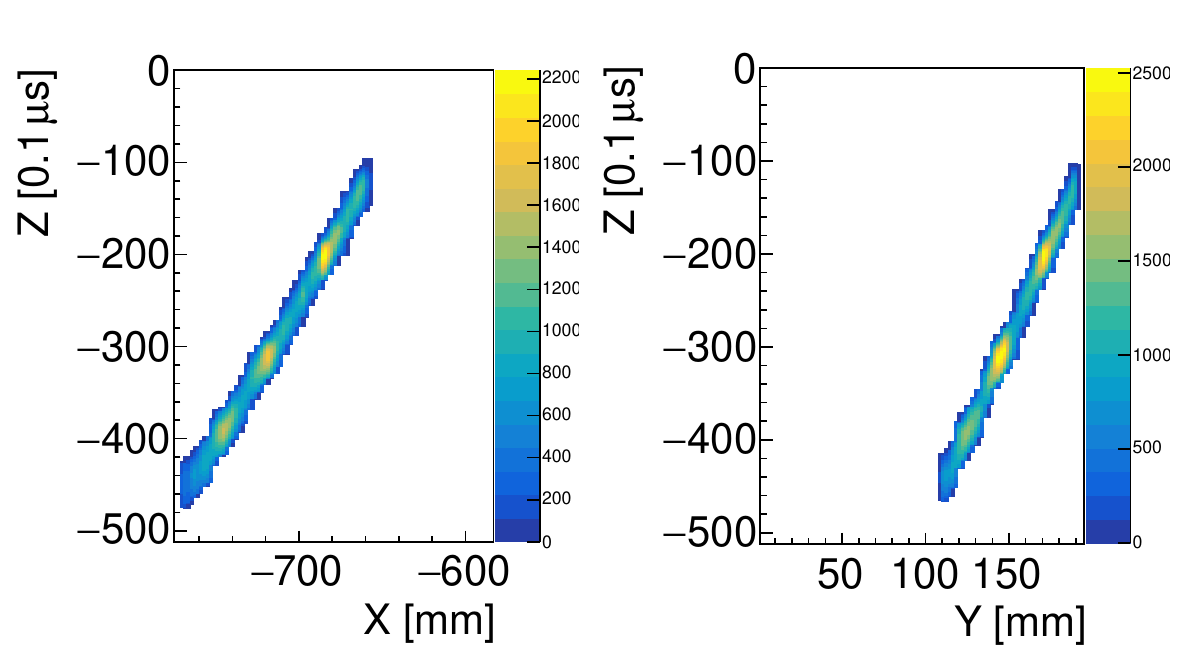}
    \caption{\label{fig:muonTrack_fullTPC} Pulses collected on each triggered strip (Left) and projected tracks (Right) of a cosmic-ray muon event detected.}
\end{figure}

All components of the detector have been assembled at SJTU as shown in Fig.~\ref{fig:fullTPC_assembly}, including the vessel, the electric field shaping cage, and the charge readout plane.
%The crane is used to lift the top flange and the electric field shaping cage.
We are testing the detector with 1 bar Ar-(5${\rm \%}$)Isobutane gas mixtures.
Data were taken with cosmic-ray muons and a typical muon event is shown in Fig.~\ref{fig:muonTrack_fullTPC}. 
The commissioning and test are ongoing to check the detector performance such as the gain uniformity, and the functional confirmation of all the sub-systems such as the electronics and gas handling systems.

\section{Micromegas tests in the PandaX-III prototype TPC}
\label{sec:prototype}
\begin{figure}[htbp]
    \centering
    \includegraphics[width=.40\textwidth]{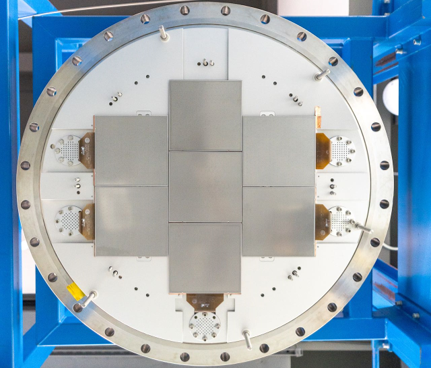}
    \qquad
    \includegraphics[width=.40\textwidth]{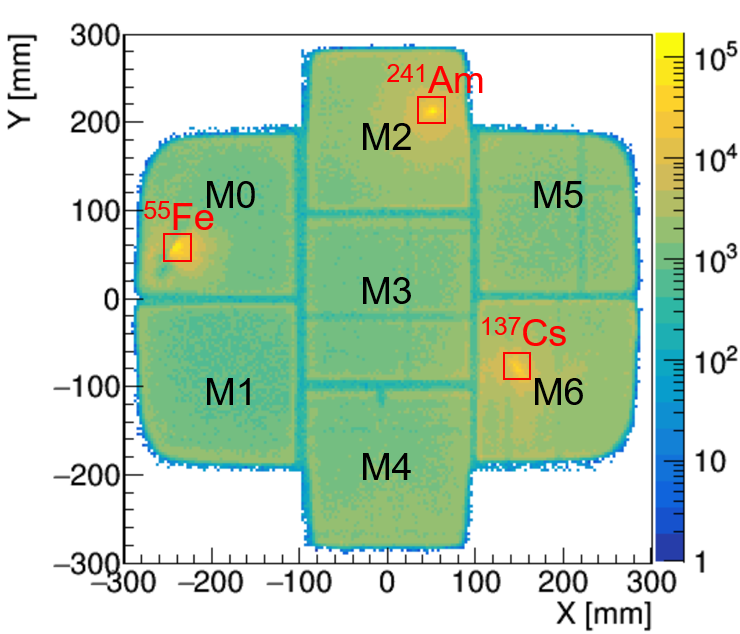}
    \caption{\label{fig:readoutPlane_hitmap_prototypeTPC} (Left) The picture of the charge readout plane with seven Micromegas modules. (Right) The hitmap of the events detected in 1 bar Ar-(3.5${\rm \%}$)Isobutane.}
\end{figure}
 The Micromegas performance is tested with the PandaX-III prototype detector. 
Its detailed design can be found in~\cite{Lin:2018mpd, Zhang:2023hsa}.
It consists of a 600~L total inner volume with a charge readout plane on the top, a cathode at the bottom, and an electric field shaping cage in the middle. 
Seven ${\rm 20.0\times20.0~cm^2}$ thermal-bonding Micromegas modules have been mounted on the charge readout plane as shown in Fig.~\ref{fig:readoutPlane_hitmap_prototypeTPC} (Left).
The electric field shaping cage has a height of 78.0~cm. The test was performed in argon-based gas mixtures at different pressures (1 and 10 bar).

\begin{figure}[tbp]
    \centering
    \includegraphics[width=.325\textwidth]{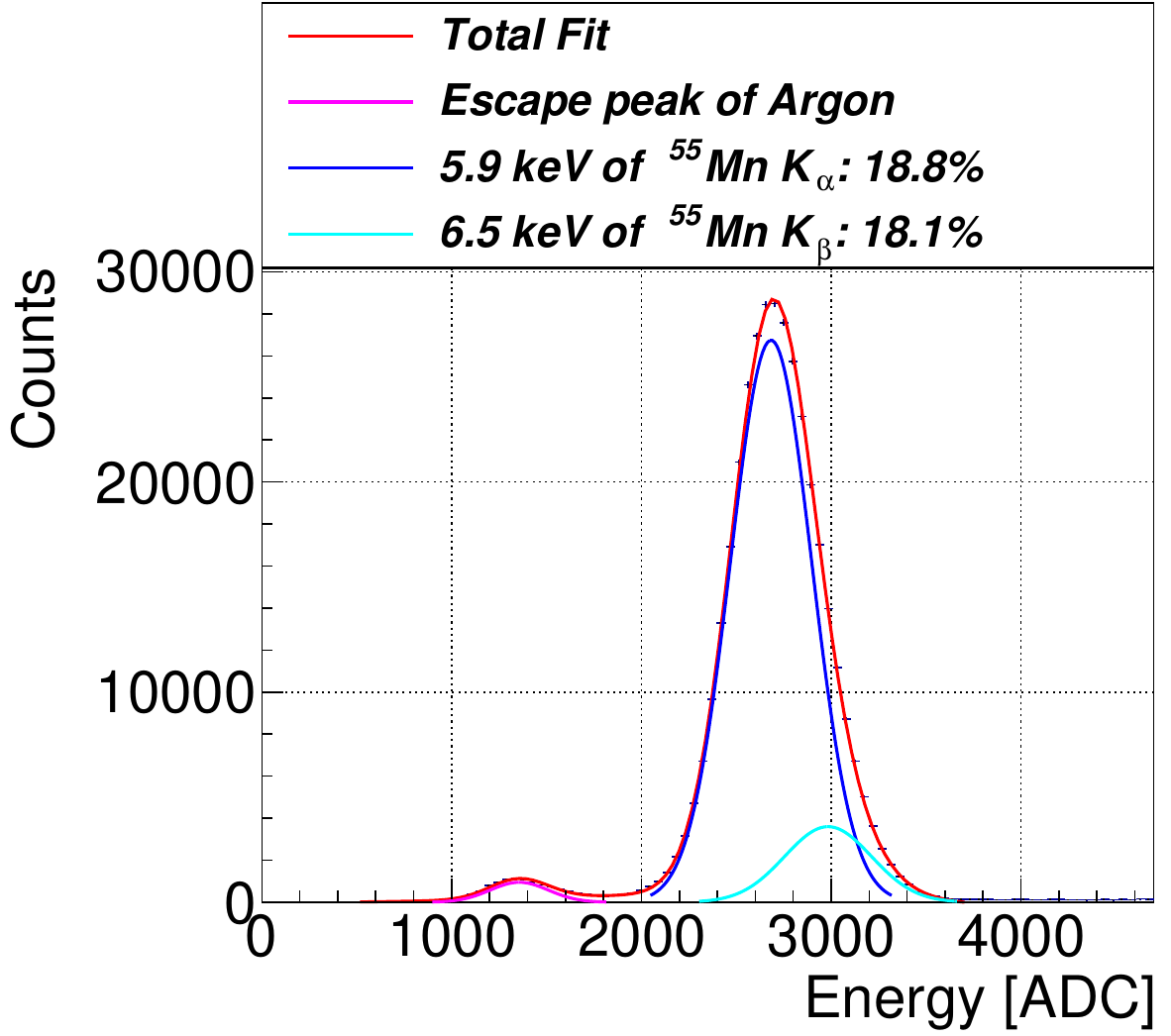}
    \includegraphics[width=.325\textwidth]{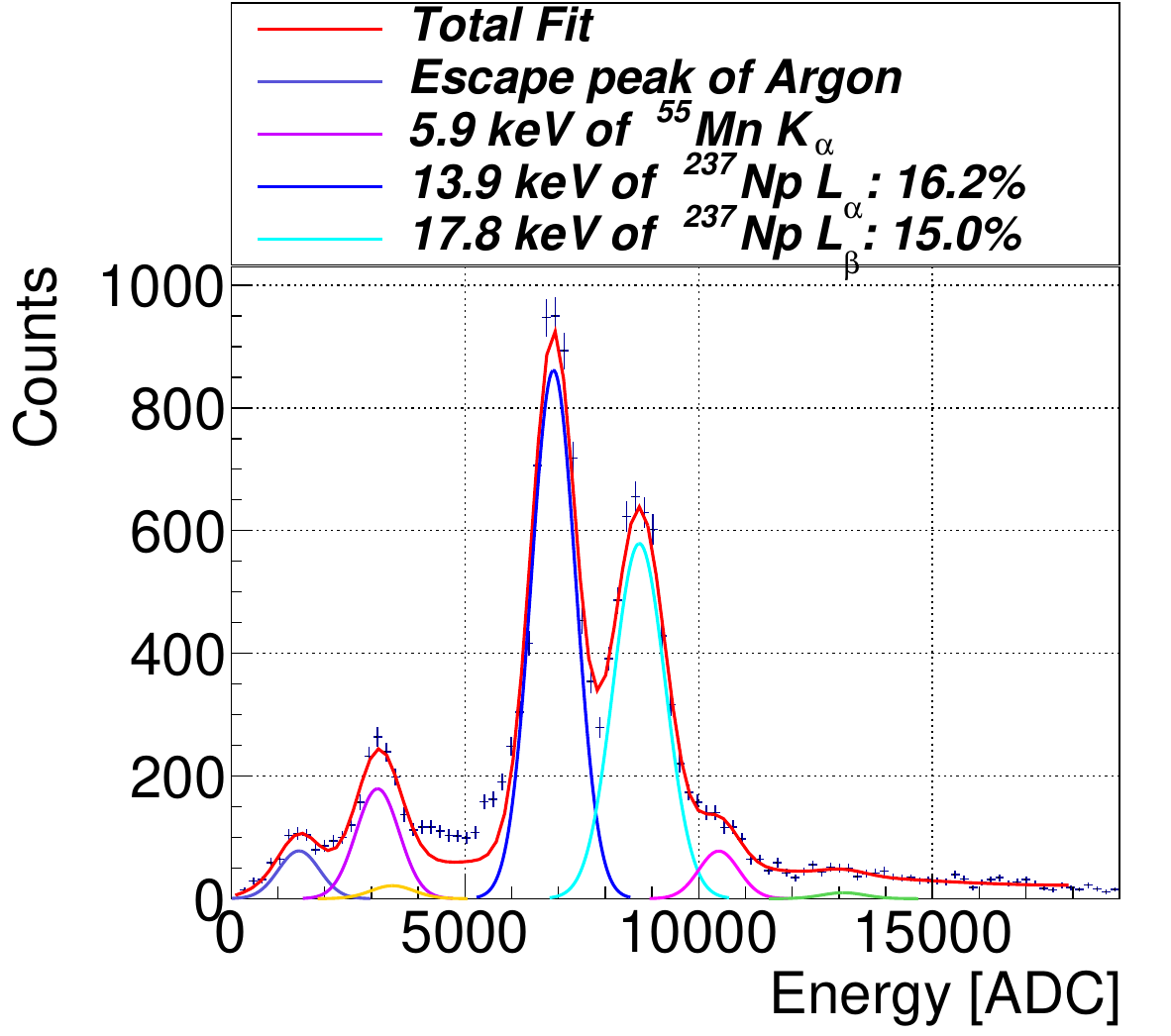}
    \includegraphics[width=.325\textwidth]{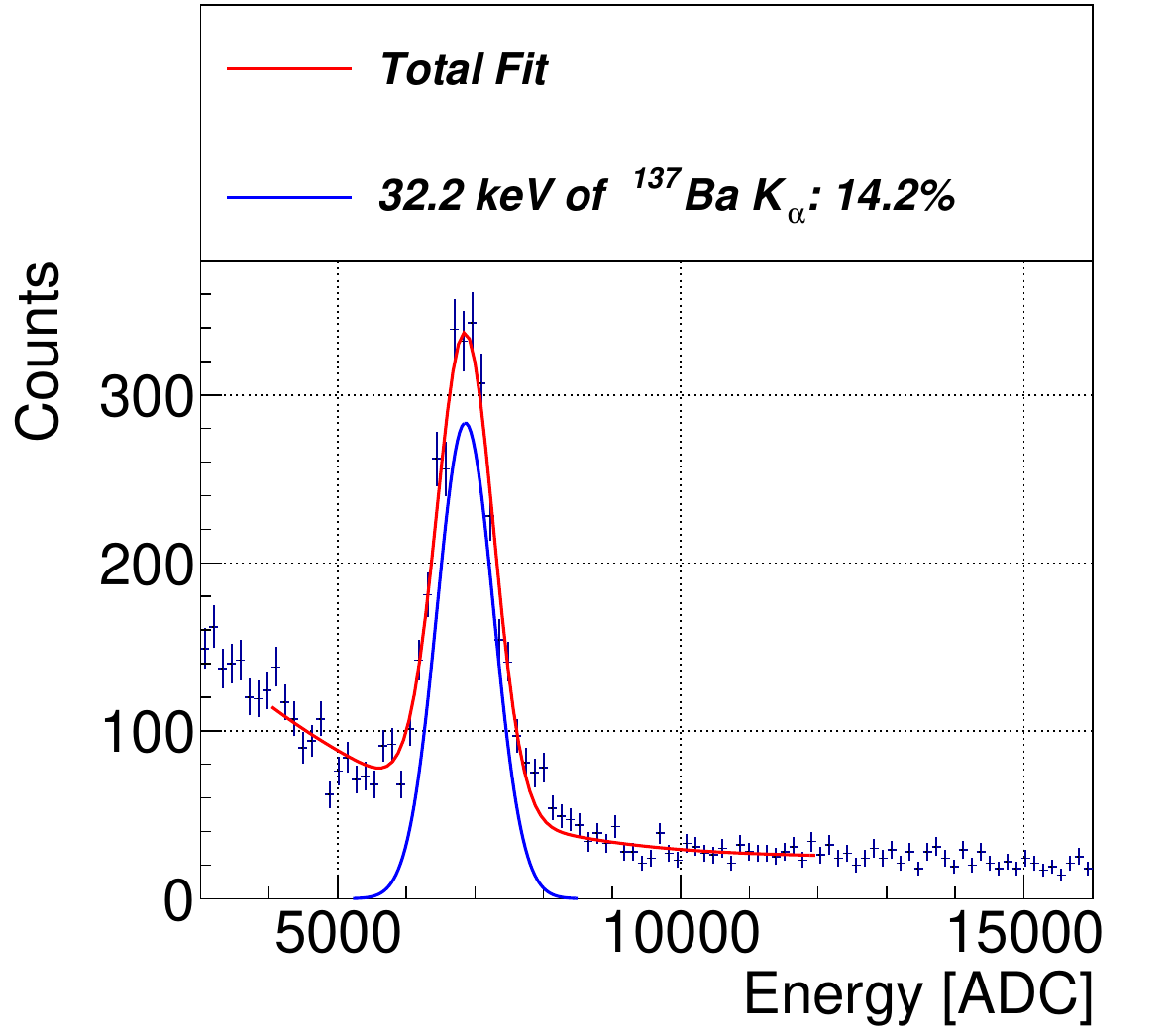}
    \caption{\label{fig:spe_1bar} 
    The energy spectra in 1 bar Ar-(3.5${\rm \%}$)Isobutane. 
    (Left) Spectrum of ${\rm ^{55}Fe}$. 
    Both the ${\rm K_\alpha}$ and ${\rm K_\beta}$ peak of ${\rm ^{55}Mn}$ are depicted at 5.9 keV and 6.5 keV.
    The escape peaks are located at 2.9 keV with an absorption energy of 3.0 keV for argon.
    (Middle) Spectrum of ${\rm ^{241}Am}$. 
    The ${\rm L_\alpha}$ and ${\rm L_\beta}$ of ${\rm ^{237}Np}$ show prominent peaks at 13.9 keV and 17.8 keV.
    (Right) Spectrum of ${\rm ^{137}Cs}$. 
    The 32.2 keV X-ray emission line of ${\rm ^{137}Ba}$ source is depicted.
    }
\end{figure}
\subsection{Performances in normal pressure}
\begin{figure}[htbp]
    \centering
    \includegraphics[width=1.0\textwidth]{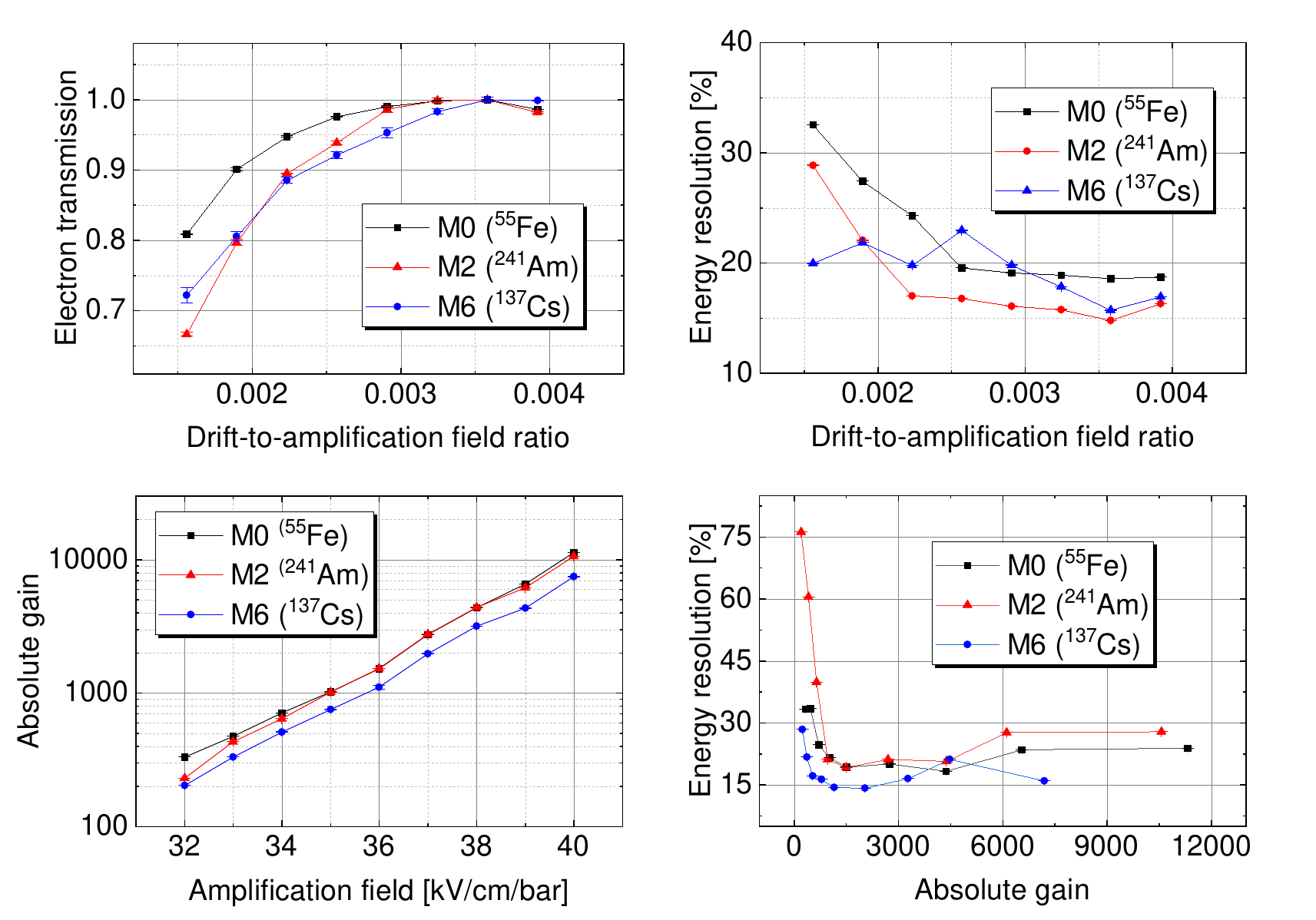}
    \caption{\label{fig:data_1bar}
    Performance tests in 1 bar Ar-(3.5${\rm \%}$)Isobutane.
    The electron transmission~(Top-left) and the energy resolution~(Top-right) evolve with the drift-to-amplification field. 
    The gain has been normalized to the maximum of each series.
    The absolute gain~(Bottom-left) and the energy resolution~(Bottom-right) evolve with the amplification field.}
\end{figure}

The detector was first tested in 1 bar Ar-(3.5${\rm \%}$)Isobutane gas mixtures. Three radioactive sources are used for energy calibration inside the TPC: a ${\rm ^{241}Am}$ source and a ${\rm ^{137}Cs}$ source are placed on the cathode, and a ${\rm ^{55}Fe}$ source is suspended 5.0~cm below the readout plane. 
The sources are wrapped in Kapton to shield the ${\alpha}$ and ${\beta}$ particles. Fig.~\ref{fig:readoutPlane_hitmap_prototypeTPC}~(Right) shows the hitmap of the events detected. 
The hot spots in the red box show the image of the sources.
The seven Micromegas are named from M0 to M6 for the convenience of calibration.
%X-ray source events in the red square of ${\rm 25 \times 25~mm^2}$ are selected by the positioning cut to calibrate the detector gain and energy resolution of corresponding Micromegas.
The calibration spectra are shown in Fig.~\ref{fig:spe_1bar}.
The spectra are fitted with a multi-Gaussian plus a polynomial function to eliminate the linear background. The events from the sources are selected by positioning cuts of a square of ${\rm 25 \times 25~mm^2}$, and the energy is reconstructed: 5.9~keV from ${\rm ^{55}Fe}$, 13.9~keV and 17.8~keV from ${\rm ^{241}Am}$, and 32.2 keV from ${\rm ^{137}Cs}$. 

The dependence of electron transmission and energy resolution with the drift-to-amplification field is tested. In a TPC, the probability of primary electrons passing from the drift volume to the amplification gap through the mesh holes of Micromegas is characterized by electron transmission. 
It includes two mechanisms: the electron attachment and recombination in the drift volume, and the transparency of the mesh electrode. 
The drift voltage was varied in the test at a fixed mesh voltage of -380~V to obtain the dependence of the electron transmission with the drift-to-amplification field ratio, as shown in Fig.~\ref{fig:data_1bar} (Top-left). 
At very low drift fields, the electron transmission is reduced by electron attachment and recombination of primary electrons generated in the drift volume. 
The transmission plateau appears at high drift fields.
The dependence of the energy resolution with the drift-to-amplification field ratio is shown in Fig.~\ref{fig:data_1bar} (Top-right). 
Basically, the resolution decreases with the electron transmission and achieves the best level at the plateau.

The dependence of gain and energy resolution with the amplification field is tested. The mesh voltage is varied at a fixed drift voltage of -12~kV.
The gain curves on the amplification field are shown in Fig.~\ref{fig:data_1bar} (Bottom-left). %which is defined as the ratio of the total charges after the avalanche and the deposited charges of primary electrons.
The average ionization energy is taken as 26.2 eV for argon-based gas~\cite{Smirnov:2005yi}.
The gain curves show a linear behavior with the amplification field in the semi-log plot and achieve maximum gains of up to over ten thousand.
Due to a slight difference in the size of the amplification gap generated during the manufacturing process, different Micromegas exhibit different performances in terms of gain with the same amplification field. 
The energy resolution evolving with the amplification field, more specifically with the gain is shown in Fig.~\ref{fig:data_1bar} (Bottom-right).
The optimal energy resolution was obtained at a gain of about 2000.
At low gains, the resolution degrades because the signal becomes comparable to the electronic noise. 
At high gains, the increase in avalanche fluctuations also leads to a degradation of the resolution~\cite{Schindler:2010los}.
The optimal energy resolution is obtained for each Micromegas: 18.3\% FWHM at 5.9 keV for M0, 14.8\% at 13.9 keV for M2, and 14.2\% at 32.2 keV for M6.

\subsection{Performances in high pressure}
\begin{figure}[htbp]
    \centering
    \includegraphics[width=1.0\textwidth]{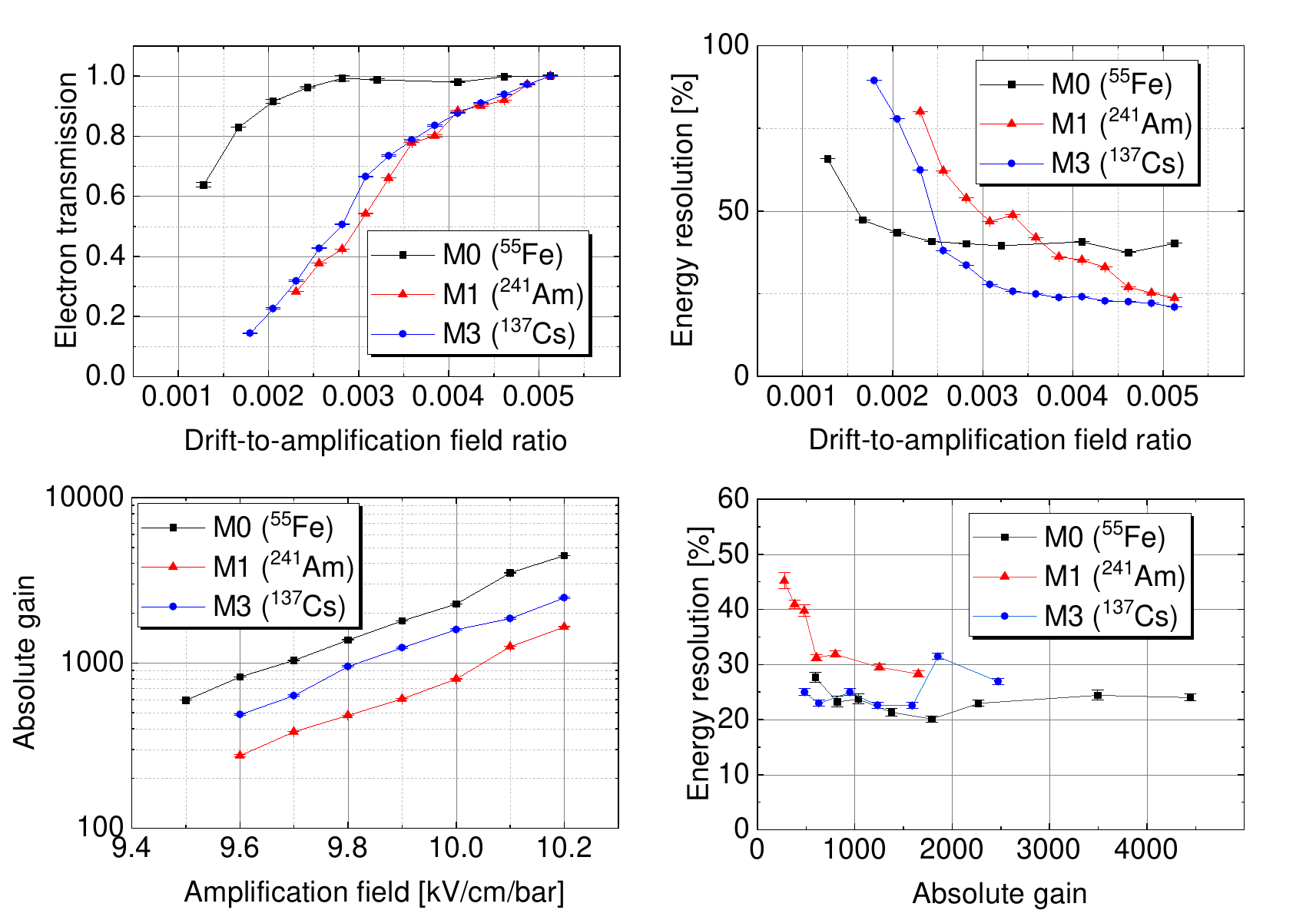}
    \caption{\label{fig:data_10bar}
    Performance tests in 10 bar Ar-(2.5${\rm \%}$)Isobutane.
    The electron transmission~(Top-left) and the energy resolution~(Top-right) evolve with the drift-to-amplification field. 
    The gain has been normalized to the maximum of each series.
    The absolute gain~(Bottom-left) and the energy resolution~(Bottom-right) evolve with the amplification field.}
\end{figure}

In high-pressure testing runs, we used 10 bar Ar-(2.5${\rm \%}$)Isobutane gas mixtures.
Besides, We have moved ${\rm ^{241}Am}$ source and ${\rm ^{137}Cs}$ source on the cathode to calibrate different Micromegas: ${\rm ^{55}Fe}$ still for M0, ${\rm ^{241}Am}$ for M1, and ${\rm ^{137}Cs}$ for M3. 
The electron transmission curve is shown in Fig.~\ref{fig:data_10bar} (Top-left), and the plateau is hard to reach in 10-bar gas for M1 and M3 due to a long drift distance of 78.0 cm. 
However, there is a long plateau for M0 with a suspended ${\rm ^{55}Fe}$ source owing to a short drift distance of about 5.0 cm. 
The dependence of the energy resolution with the drift-to-amplification field ratio is shown in Fig.~\ref{fig:data_10bar} (Top-right).
The gain curves in Fig.~\ref{fig:data_10bar} (Bottom-left) show that the maximum gain obtained in 10 bar is up to several thousand.
The energy resolution evolving with the gain is shown in Fig.~\ref{fig:data_10bar} (Bottom-right).
We also obtained the optimal energy resolution for each Micromegas: 20.1\% FWHM at 5.9 keV for M0, 23.7\% at 13.9 keV for M1, and 22.5\% at 32.2 keV for M3. 
The optimal energy resolution at 10 bar did not attain the result of the 1 bar test.

\section{Conclusions and Outlook}
\label{sec:outlook}
The PandaX-III uses a high-pressure xenon gaseous TPC to search for NLDBD of $^{136}$Xe. 
The gaseous TPC records both energy depositions and tracks of an event with unique background suppression capability. A prototype TPC with 7 Micromegas modules has been built to demonstrate the design and the performance of Micromegas-based TPC. The key indicators of TPC, including the electron transmission, gain, and energy resolution, were tested with three radioactive sources inside (${\rm ^{241}Am}$, ${\rm ^{137}Cs}$, and ${\rm ^{55}Fe}$). The maximum gain achieved is up to several thousand in argon-based gas mixtures.
The optimal energy resolution reached 20.1\% FWHM at 5.9 keV at a high pressure of 10 bar.
%in our test in 1 bar: 18.3\% FWHM at 5.9 keV for M0, 14.8\% at 13.9 keV for M2, and 14.2\% at 32.2 keV for M6 (in 10 bar: 20.1\% FWHM at 5.9 keV for M0, 23.7\% at 13.9 keV for M1, and 22.5\% at 32.2 keV for M3).

The design and current status of the core components of the detector are presented in this paper. The construction of the full-size PandaX-III detector has been completed. 
The vessel remained stable for over one month at 10 bar pressure. 
The drift electric field can reach the baseline design of 1~kV/cm.
52 Micromegas have been mounted on the charge readout plane. All the components of the detector were assembled. 
We are testing the detector with 1 bar argon-based gas mixtures. Performance tests in high-pressure xenon gas will be further performed in the future. 

\acknowledgments
{This work was supported by grants No.U1965201 and No.11905127 from the National Natural Sciences Foundation of China and grant 2016YFA0400302 from the Ministry of Science and Technology of China. We thank the support from the Key Laboratory for Particle Physics, Astrophysics and Cosmology, Ministry of Education. We thank the support from
the Double First Class Plan of Shanghai Jiao Tong University.}

\bibliographystyle{unsrt}
\bibliography{Status_PandaX-III}
\end{document}